\documentstyle[12pt]{article}
\makeatletter
\newcommand{\singlespacing}{\let\CS=\@currsize\renewcommand{\baselinestretch}
{1.0}\tiny\CS}
\newcommand{\doublespacing}{\let\CS=\@currsize\renewcommand{\baselinestretch}
{1.5}\tiny\CS}

\newcommand {\beq} {\begin{equation}}
\newcommand {\al} {\alpha}
\newcommand {\eeq} {\end{equation}}
\newcommand {\bt} {\beta}
\newcommand {\om} {\omega}
\newcommand {\dl} {\delta}
\newcommand {\Dl} {\Delta}
\date{}
\begin{document}

\thispagestyle{empty}\setcounter{page}{1}
\vskip10pt
\centerline{\bf BROADENING OF SPECTRAL LINES DUE TO
 DYNAMIC MULTIPLE
SCATTERING}
\centerline{ \bf AND THE TULLY-FISHER RELATION}

\vskip20pt
\centerline{\footnotesize Sisir Roy$^{1,2}$, \& 
Menas Kafatos$^1$  }
\centerline{\footnotesize \it Center for Earth Observing and Space Research}
\centerline{\footnotesize \it Institute for Computational Sciences and Informatics and}
\centerline{\footnotesize \it Department of Physics, George Mason University} 
 \baselineskip=10pt
\centerline{\footnotesize \it Fairfax,  VA  22030  USA}
\vskip10pt
\centerline{\footnotesize Suman Dutta$^2$ }
\centerline{\footnotesize \it Physics and Applied Mathematics Unit}
\centerline{\footnotesize \it Indian Statistical Institute, Calcutta, INDIA}
\vskip5pt
\centerline{\footnotesize $^{1,2}$ e.mail: sroy2@osf1.gmu.edu  }
\centerline{\footnotesize $^1$ e.mail : mkafatos@compton.gmu.edu}
\centerline{\footnotesize $^2$ email : res9428@isical.ac.in }

\vskip20pt
\doublespacing
\abstract{\noindent{\small{  The frequency shift of spectral lines
is most often explained by the Doppler Effect in terms of relative motion, whereas 
 theDoppler broadening
of a particular line mainly depends on the absolute temperature. The Wolf effect on the
other hand deals with the correlation induced spectral change and explains
both the broadening and shift of the spectral lines. In this framework
a relation between the width of the spectral line is related to the
redshift $z$ for the line and hence with the distance. For smaller
values of $ z$ a relation similar to the Tully-Fisher relation can be obtained and 
for larger
values of $z$  a more general relation can be constructed. The derivation of
this kind of relation based on dynamic multiple scattering theory may play
a significant role in explaining the overall spectra of quasi stellar objects.
We emphasize that this mechanism is not applicable for nearby galaxies,
 $z \leq 1$.

\newpage

\section{{\bf Introduction}}

In studying the motion of astronomical objects, astrophysicists utilize the 
study of frequency  shift of spectral lines.
In general, an emphasis on relating the shift of a spectral line
with its width has not
been made so far. The Tully-Fisher relation$^{(1)}$ is
an empirical correlation which finds that the luminosity $L$ of a disk galaxy is
proportional to its maxium rotational velocity $ V^\alpha _{max}$ , where
$\alpha$ has been observationally established to be ~(3 -4)$^{(2)}$. In spite
of the frequenty use of the TF relation as a distance indicator, the
physical origin of this relationship is poorly understood , and it remains unclear
whether all rotationally supported disk galaxies , including late-type spirals
and irregulars, obey a single luminosity-line width correlation.
Recently, Matthews et al $^{(3)}$ made an attempt to analyze this situation.
It is generally argued that in order to obtain a measure of the maximum
rotational velocity of a disk from its measured global $H-I$ profile width ,
some correction to the observed line width should be made for the effects
of broadening due to turbulent (i.e. non-rotational) motions
( cf. Roberts$^{(4)}$, Bottinelli et al. $^{(5)}$) .
A linear summation of rotational and random motions adequately describes the
observed profile widths of giant galaxies , while for dwarfs galaxies( i.e. slowly
rotating galaxies with Gaussian-like line profiles ) , a sum in the quadrature
of the random and rotational terms is appropriate. However, Rhee$^{(6)}$
has pointed out the shortcomings of this type of addition for the type
of objects considered here.

On the other hand, a dynamic multiple scattering theory$^{(7),(8)}$ derived from the
field of Statistical Optics has been
developed to account for the shift of a spectral line as well as the
broadening of the line. It
is shown that when light passes through a turbulent(or inhomogeneous ) medium,
due to multiple scattering the shift and the width can be calculated.
Here, a sufficient condition for redshift has been derived and when applicable
 the shift is shown to be larger than broadening. The width of the
spectral line can be calculated after multiple scatterings and a relation
can be derived between the width and the shift $z$ which applies to
any value of $z$. 
 For small values of $z$ this can be reduced to a Tully-Fisher
type of relation. It would be interesting to estimate this width and make a
comparison with the correction part of the line profile width considerd
by Bottinelli et al$^{(5)}$ applying to non-rotational motion. We want to emphasize 
that it is then possible to derive a relation for
distance indicator ( which reduces to a Tully-Fisher type relation as a special case ) 
by studying
the particular collisional mechanism which itself is  a physical process. 
 This kind of physical process can be identified even in the laboratory
experiments and as such it is plausible that it applies to astronomical objects.

In this paper we shall start with a brief discussion of the various type of 
broadenings  (section 2). In section 3  we shall discuss the main results of our
 Multiple Scattering Theory within the Wolf framework of redshifts$^{(9)}$. 
Based on this approach a relation between the shfit $z$ and the width will
be derived in section 4. Finally, the possible implications for QSO observations will
be discussed in section 5.

\section{\bf Spectral Broadening}

If a spectral line is examined by means of a spectrograph , its width is
dependent on the slit width employed. In general, the narrower the slit width
 the less broad is the resulting spectral line as recorded on a  photographic
plate. Nevertheless, however narrow the slit width is , the sharpest line has a
finite width even for the best optical system. Three causes producing
the breadth of a spectral line are :

\begin{enumerate}

\item Its natural width

\vskip5pt
\item Doppler effect
\vskip5pt
\item external effect
\vskip5pt
\end{enumerate}
\vskip10pt
\subsection{Natural Line Width}

An atom can stay in two type of states. A state in which an atom free from
external effects can remain for an arbitrarily long time is called a stationary
state. Only the ground state is stationary since it corresponds to the
minimum possible energy for a given atom. An excited state is non-stationary
since spontaneous transitions to lower energy levels are possible for it.
Such states are called transient states which are characterized by a finite life 
time. A quantitative measure of instability of an excited state
is the time $\tau$ during which the number of atoms in a system at
a given excited state decreases by a factor of e . This quantity $\tau$
is known as the {\it lifetime of the excited state} and coincides with
the average time spent by the atoms in the excited state. In quantum mechanics,
$\tau$ is associated with the probability of spontaneous radiative transitions
from a given excited state to a lower energy level. Life times of 
excited states of atoms generally lie in the range $10^{-9}$ to $10^{-8}$
second.
The finiteness of the life time of an atom in a transient state can easily
be taken into account by introducing the damping factor into the expression for the
wave function
\begin{equation}
\psi(\vec {r} , t) = \exp ^{[-\frac{i}{h}E t - \frac{\gamma t}{2}]}
\end{equation}

where $\gamma$ is a positive constant. It should be noted that the wave function
in stationary state is
\begin{equation}
\psi(\vec {r} , t) = \exp ^{[-\frac{i}{h}E t ]}\psi(\vec{r})
\end{equation}

Therefore the above function not only oscillates with a frequency $\omega =
\frac{E}{h}$, but also attenuates with time due to the presence of the factor
$\exp^{[- \frac{\gamma t}{2}]}$. The probability density of finding the atom
in the state $\psi$ is given by
\begin{equation}
{|\psi(\vec r,t)|}^2 = \exp{[-\gamma t]} {| \psi(\vec r)|}^2
\end{equation}

The time during which the probability density diminishes  to $\frac{1}{e}$ th
 of its initial value is obviously the life time $\tau$ of the state.
It then follows that
\begin{equation}
\gamma \tau = 1
\end{equation}

The quantity $\gamma$ defined by this expression is called the 
{\it damping constant}.

Since energy and time are canonically conjugate quantities, according to
the Heisenberg Uncertainty relation, the energy of an excited state is not exactly
definite. The indeterminacy $\Gamma$ in the energy of a transient state is
connected with its lifetime $\tau $ through the relation
\begin{equation}
\Gamma \sim \frac{\hbar}{\tau} = \hbar \gamma 
\end{equation}

Blurring of  adjacent energy levels can generally be due to various reasons. The quantity
$\Gamma$ associated with the probability of spontaneous radiative transitions is called the
natural width of the level. If an atom remains in the normal state for a long time
the uncertainty of the energy value is small and the level is sharp. If the electron
is excited to an upper level where it remains for some time, the
uncertainty of the energy value is greater and the width of the level
is greater. The natural
 line width is the minimum limit for the radiation line width.

\subsection{Doppler Broadening}

The thermal motion of emitting atoms leads to the so-called Doppler broadening
of spectral lines. Owing to thermal agitation, most of the atoms emitting light have high velocities.
The random motion of the atoms and the molecules in a gas, however, produce
 a net broadening of the lines with no apparent shift in its central maximum. The
frequency spread of this line is called the half-intensity breadth, and 
is given by
\begin{equation}
\delta = 1.67 {\frac{\nu_0}{c}} \sqrt{\frac{2 R T}{m}}
\end{equation}
\vskip5pt
\noindent
The half intensity breadth is defined as the interval between two points where
the intensity drops to half its maximum vale. Here, $R$ is the  gas
constant, $T$ is the  temperature of the gas in K, and $m$ is the atomic
weight. Therefore the Doppler broadening is

\begin{enumerate}
\item proportional to the frequency $\nu_0$

\item proportional to the square root of the (absolute) temperature $T$; and

\item inversely proportional to the square root of the atomic weight $m$.
\end{enumerate}
\vskip5pt
\noindent
Experimental observations indicate that in keeping with the above equation, in order 
to produce
sharper lines in any given spectrum, the temperature must be lowered. Furthermore, for
a given temperature 
the lines produced by the lighter elements in the periodic table are in general
broader than those produced by the heavy elements.
\vskip5pt
\noindent
A comparison of the estimated values of natural width and Doppler width shows
 that the Doppler width at room temperatures is much larger than the natural
width [ Figure 1].

\subsection{Collisional Broadening}

The  spectral broadenings described in the previous two subsections,
can be classified as due to internal effects. There exist other reasons for
spectral  line broadening specifically an excited atom in a gas with a finite number density
of particles undergoes collisions with neighbouring atoms. Since the phase of
radiation changes upon each collision, the monochromatic nature of the emission
line is violated. This is effectively taken into account by introducing the total
level width, which is equal to $ \Gamma + \Gamma_{\rm {col}}$, where
$\Gamma_{\rm {col}} = \frac{1}{\pi \tau_0}$ is the collision level width ( $\tau_0$ is
the mean free time of the atom in gaseous medium).
Another mechanism, which can explain the line broadening, is an application of statistical
optics where the existence of a random dielectric susceptibility which fluctuates
both spatially and temporally is assumed. The scattering of light induced by the random susceptibilty
can produce both a shift and a broadening of the spectral line. This is called the 
Wolf effect$^{(9)}$,
the main features of which are described as follows.
\vskip10pt
\section{{\bf Dynamic Multiple Scattering Theory}}

First, we briefly state the main results of Wolf's scattering mechanism. Let us
consider a polychromatic electromagnetic field of light of central frequency $\omega_0$ 
and 
width $\delta_0$, incident on the scatterer. The incident spectrum is assumed to be of the
form
\begin{equation}\displaystyle{
S^{(i)}(\omega)= A_0e^{\left[-\frac{1}{2\delta_0^2}(\omega-\omega_0)^2\right]}
}
\end{equation}
The spectrum of the scattered field is given by$^{(10)}$
\begin{equation}\displaystyle{
S^{(\infty)}(r \vec{u'} ,\omega')=A\omega'^4\int_{-\infty}^{\infty}
K(\omega,\omega',\vec{u}, \vec{u'})S^{(i)}(\omega)d\omega}
\end{equation}
which is valid within the first order Born approximation$^{(11)}$. Here $K(\omega, \omega')$ is the so called scattering kernel and it
plays the most important role in this mechanism. $\vec{u}$ and $\vec{u'}$ are the
unit vectors in the direction of incident and scattered fields respectively.
Instead of studying $\cal K(\omega,\omega')$ in detail, we consider a particular case for the
correlation function $G(\vec{R},T;\omega) $ of the generalized dielectric
susceptibility $\eta(\vec{r},t;\omega)$ of the medium which is characterized by an
anisotropic Gaussian function
\begin{equation}\displaystyle{
\begin{array}{lcl}
G(\vec{R},T;\omega) & = &
<\eta^*(\vec{r}+\vec{R},t+T;\omega)\eta(\vec{r},t;\omega)>\\ \\
& = &G_0exp\left[-\frac{1}{2}\left(\frac{X^2}{\sigma_x^2}
+\frac{Y^2}{\sigma_y^2}
+\frac{Z^2}{\sigma_z^2}+\frac{c^2T^2}{\sigma_\tau^2}\right)\right]
\end{array}
}
\end{equation}
\vskip5pt
\noindent
Here $~G_0~$ is a positive constant, $~\vec{R}~=~(X,~Y,~Z)~$, and
$~\sigma_x~,~\sigma_y~,~\sigma_z~,~\sigma_\tau~$ are correlation 
lengths. The anisotropy is indicated by the unequal
correlation lengths in different spatial as well as temporal directions. $\cal
K(\omega,\omega')$ can be obtained from the four dimensional Fourier Transform
of the correlation function $G(\vec{R},T;\omega) $.  In this case $\cal
K(\omega,\omega')$ can be shown to be of the form
\begin{equation}\displaystyle{
{\cal{K}} (\omega,\omega')=B exp\left[-{\frac{1}{2}} \left(\alpha'\omega'^2-
2\beta\omega\omega'+\alpha\omega^2 \right) \right]
}
\end{equation}
where
\begin{equation}\displaystyle{
\left.
\begin{array}{lcl}
\alpha &=& {\frac{\sigma_x^2}{c^2}}u_x^2+{\frac{\sigma_y^2}{c^2}}u_y^2+
{\frac{\sigma_z^2}{c^2}}u_z^2+{\frac{\sigma_\tau^2}{c^2}}\\ \\
\alpha' &=& {\frac{\sigma_x^2}{c^2}}u_x'^2+{\frac{\sigma_y^2}{c^2}}u_y'^2+
{\frac{\sigma_z^2}{c^2}}u_z'^2+{\frac{\sigma_\tau^2}{c^2}}\\ \\
{\rm and} \ \ \beta &=& {\frac{\sigma_x^2}{c^2}}u_xu_x'+{\frac{\sigma_y^2}{c^2}}u_yu_y'+
{\frac{\sigma_z^2}{c^2}}u_zu_z'+{\frac{\sigma_\tau^2}{c^2}}
\end{array}
\right\}
}
\end{equation}
Here $\hat{u}=(u_x,u_y,u_z)$ and $ \hat{u'}=(u_x',u_y',u_z')$ are the unit vectors in the 
directions of the incident and scattered fields respectively.

Substituting (1) \& (4) in (2), we finally get 
\begin{equation}\displaystyle{ S^{(\infty)}(\omega')=
A'e^{\left[-\frac{1}{2\delta_0'^2}(\omega'-\bar{\omega}_0 )^2\right]} }
\end{equation}
where
\begin{equation}\displaystyle{
\left.
\begin{array}{lcl}
\bar{\omega}_0 &=& \frac{|\beta|\omega_0}{\alpha'+\delta_0^2
(\alpha\alpha'-\beta^2)}\\ \\
\delta_0'^2 &=& \frac{\alpha\delta_0^2+1}{\alpha'+\delta_0^2
(\alpha\alpha'-\beta^2)}\\ \\
 {\rm and} \ \ A' &=& \sqrt{\frac{\pi}{2(\alpha\delta_0^2+1)}}ABA_0\omega_0'^4\delta_0
 exp\left[\frac{|\beta|\omega_0\bar{\omega}_0-\alpha\omega_0^2}
{2(\alpha\delta_0^2+1)}\right]
\end{array}
\right\}
}
\end{equation}
Though $A'$ depends on $\omega'$, it was approximated by James and
Wolf$^{(8)}$ 
to be a constant so that $S^{(\infty)}(\omega')$ can be considered to be Gaussian.

The relative frequency shift is defined as 
\begin{equation}\displaystyle{
z=\frac{\omega_0-\bar{\omega}_0}{ \bar{\omega}_0}
}
\end{equation}
where $\omega_0$ and $\bar{\omega}_0$ denote the  unshifted and shifted central frequencies 
respectively. We say that the spectrum is redshifted or blueshifted 
according to whether $~z~>~0~ ~
or~~z~<~0~$ respectively. Here
\begin{equation}
\displaystyle{
z=\frac{\alpha'+\delta_0^2(\alpha\alpha'-\beta^2)}{|\beta|}-1
}
\end{equation}
\vskip5pt
\noindent
It is important to note that this $z$-number does not depend on the incident frequency, 
$\omega_0$. This is a very important aspect if the mechanism is 
to apply in the 
astronomical domain. Expression (15) implies that the spectrum can be shifted to the blue 
or to the red, according to the sign of the term $\alpha'+\delta_0^2(\alpha\alpha'-\beta^2)~>~
|\beta|~$. To obtain the no-blueshift condition, we use Schwarz's Inequality
which implies that $\alpha\alpha'-\beta^2~\geq~0~$. Thus, we can take 
$$\displaystyle{
 \alpha'~>~|\beta|
 }$$
as the sufficient condition to have only redshift by that mechanism.

Let's now assume that the light in its journey encounters many such scatterers. What we 
observe at the end is the light scattered many times, with an effect as that stated 
above in every individual process. Let there be N scatterers  between the source and the
observer and $z_n$ denote the relative frequency shift after the $n^{th}$ scattering of the 
incident light from the $(n-1)^{th}$ scatterer, with $\omega_n$ and $\omega_{n-1}$ being the 
central frequencies of the incident spectra at $n^{th}$ and $(n-1)^{th}$ scatterers. Then by 
definition,
$$\displaystyle{
 z_n =\frac{ \omega_{n-1}-\omega_n}{\omega_n}, ~~~~~~n~=~1,~2,~.~.~.~.,~N
 }$$
or,
$$\displaystyle{
 \frac{\omega_{n-1}}{\omega_n}=1+z_n,~~~~~~n~=~1,~2,~.~.~.~.,~N
 }$$ 
Taking the product over $ n$ from $n ~=~ 1$ to $n~ =~ N$, we get,
$$ \displaystyle{
 \frac{\omega_0}{\omega_N}=(1+z_1)(1+z_2)~.~.~.~.~.~(1+z_N)
 }$$
 The left hand side of the above equation is nothing but the ratio of the
source frequency and the final or  observed frequency $z_f$. Hence,
\begin{equation}\displaystyle{
1 + z_f = (1+z_1)(1+z_2)~.~.~.~.~.~(1+z_N)
}
\end{equation}
\vskip5pt
\noindent
Since the $z$-number due to such effect does not depend upon the central
frequency of the incident spectrum, each $z_i$ depends on $\delta_{i-1}$ only,
not $\omega_{i-1}$ [here $\omega_j$ and $\delta_j$ denote the central frequency
and the width of the incident spectrum at $(j+1)^{th }$ scatterer]. To find the
exact dependence we first calculate the broadening of the spectrum after N number
of scatterings.
\vskip8pt
\subsection{Effect of Multiple Scatterings on the Spectral Line Width}
\vskip5pt
From the second equation in (13), we can easily write,
\begin{equation}\displaystyle{
\left.
\begin{array}{lcl}
\delta_{n+1}^2 & = &
\frac{\alpha\delta_n^2+1}{\alpha'+(\alpha\alpha'-\beta^2)\delta_n^2} \\ \\
& = & \left(\frac{\alpha\delta_n^2+1}{\alpha'}\right)\left[1+\delta_n^2\left(
\frac{\alpha\alpha'-\beta^2}{\alpha'}\right)\right]^{-1} 
\end{array}
\right\}}
\end{equation}

From (13), we can also write
\begin{equation}
\displaystyle{
\omega_{n+1} = 
\frac{\omega_n|\beta|}{\alpha'+(\alpha\alpha'-\beta^2)\delta_n^2} 
}
\end{equation}

Then from (14) \& (15), we can write
\begin{equation}
\displaystyle{
\left.
\begin{array}{rcl}
z_{n+1}&=&\frac{\omega_n-\omega_{n+1}}{\omega_{n+1}}\\ \\
&=&\frac{\alpha'+(\alpha\alpha'-\beta^2)\delta_n^2}{|\beta|}-1\\ \\
&=&\frac{\alpha'}{|\beta|}\left\{1+\left(\frac{\alpha\alpha'-\beta^2}{\alpha'}
\right)\delta_n^2\right\}-1
\end{array}
\right\}}
\end{equation}
\noindent
Let's assume that the redshift per scattering process is very small, {\it i.e.,}
$$0~<~\epsilon ~ = ~ z_{n+1}~~<<~1 $$ for all $n$.
\vskip5pt
\noindent
Then,
$$\displaystyle{
\begin{array}{lrcl}
& 1+\epsilon & = & \frac{\alpha'}{|\beta|}\left\{1+\left(\frac{\alpha\alpha'
-\beta^2}{\alpha'}\right)\delta_n^2\right\}\\ \\
{\rm or,}& (1+\epsilon)\frac{|\beta|}{\alpha'} & = & 1+\left(\frac{\alpha\alpha'-
\beta^2}{\alpha'}\right)\delta_n^2
\end{array}
}$$
In order to satisfy this condition and in order to have a redshift
 only ( or positive z),
we see that the first factor $\frac{\alpha'}{|\beta|}$ in the right term cannot much bigger than 1, and, more important, 
\begin{equation}
 \left(\frac{\alpha\alpha'-\beta^2}{\alpha'}\right)\delta_n^2~~<<~~1
\end{equation}

In that case, from (17), after neglecting higher order terms, the expression for 
$~~\delta_{n+1}^2~~$ can be well approximated as:
$$\displaystyle{
\delta_{n+1}^2  \approx
 \left(\frac{\alpha\delta_n^2+1}{\alpha'}\right)\left[1-\delta_n^2\left(
\frac{\alpha\alpha'-\beta^2}{\alpha'}\right)\right]
}$$
which, after carrying out a  simplification, gives a very important recurrence relation:
\begin{equation}\displaystyle{
\delta_{n+1}^2 =  \frac{1}{\alpha'}+\frac{\beta^2}{\alpha'^2}\delta_n^2.
}
\end{equation}

Therefore,
$$\displaystyle{
\begin{array}{lcl}
\delta_{n+1}^2 & = & \frac{1}{\alpha'}+\frac{\beta^2}{\alpha'^2}\delta_n^2 \\
\\ 
& = & \frac{1}{\alpha'}+\frac{\beta^2}{\alpha'^2}[ \frac{1}{\alpha'}
+\frac{\beta^2}{\alpha'^2}\delta_{n-1}^2] \\ \\
& = & \left(\frac{\beta^2}{\alpha'^2}\right)^2\delta_{n-1}^2
+\frac{1}{\alpha'}\left(1+\frac{\beta^2}{\alpha'^2}\right) \\ \\
& .~. & .~.~.~.~.~.~.~.~.~.~.~.~.~.~  \\ \\
& = & \left(\frac{\beta^2}{\alpha'^2}\right)^{n+1}\delta_0^2
+\frac{1}{\alpha'}\left(1+\frac{\beta^2}{\alpha'^2}+~.~.~.~.~
\frac{\beta^{2n}}{\alpha'^{2n}}\right).
\end{array}
}$$

Thus 
\begin{equation}\displaystyle{
\delta_{N+1}^2  =
\left(\frac{\beta^2}{\alpha'^2}\right)^{N+1}\delta_0^2
+\frac{1}{\alpha'}\left(1+\frac{\beta^2}{\alpha'^2}+~.~.~.~.~
\frac{\beta^{2N}}{\alpha'^{2N}}\right).
}
\end{equation}

As the number of scattering increases, the width  of the spectrum obviously
increases and the most important topic to be considered is whether this width
is below some tolerance limit or not, from the observational point of view.
There may be several measures of that tolerance limit. One of them is the {\it
Sharpness Ratio}, defined as
$$ Q=\frac{\omega_f}{\delta_f}$$
where $~\omega_f ~$ \& $~ \delta_f ~$ are the mean frequency \& the width of
the observed spectrum.

After $N$ number of scattering, this sharpness ratio, say $Q_N$, is given by
the following recurrence relation :
$$Q_{N+1}=Q_N \sqrt{\frac{\alpha'}{\alpha'+(\alpha\alpha'-\beta^2)\delta_N^2}-
\frac{1}{\alpha\delta_N^2+1}}.$$ 
It is easy to verify that the expression under the square root lies between 0 \&
1. Therefore, $Q_{N+1}~~<~~Q_N$, and the line is broadened as the scattering 
process goes on ( Figs.2 and 3). 

Under the sufficient
condition of redshift [i.e., $|\bt|~<~\al'$ ]$^{(12)}$ it
was shown that in the observed spectrum
\begin{equation}
\Dl\om_{n+1} \gg \dl_n
\end{equation}
\noindent
if the following condition holds:
\begin{equation}
\frac{\dl_n\om_0(\al\al'-\bt^2)}{\al'+(\al\al'-\bt^2)\dl_n^2} \gg 1
\end{equation}
where $\om_0$ is the source frequency.

The relation (23) signifies that the shift is more prominent than the effective
broadening so that the  spectral lines are observable and can  be analyzed. If, 
on the other hand ,the
broadening is higher than the shift of the spectral line, it will be impossible
to detect the shift from the blurred spectrum. Hence we can take  relation
(24) to be one of the conditions necessary for the observed spectrum to be
analyzable. For large $N$ (i.e., $N \rightarrow \infty $ ), the series in the second 
term of right hand side of (22) converges to a finite sum and we get

$$ \displaystyle{
\delta_{N+1}^2  =
\left(\frac{\beta^2}{\alpha'^2}\right)^{N+1}\delta_0^2
+\frac{\alpha'}{\alpha'^2 - \beta^2}.}$$

If $\delta_0$ is considered as arising out of Doppler broadenning only, we can 
estimate  $\delta_{\rm Dop}  \sim 10^9$  for $T = 10^4$ K. On the other hand, for anisotropic 
medium, we can take $\sigma_x = \sigma_y = 3.42 \times 10^{-1}, \ \sigma_z = 
8.73 \times 10^{-1}, \ \alpha' = 8.68 \times 10^{-30}, \ \alpha = 8.536 \times
 10^{-30} {\rm and} \beta = 8.607 \times 10^{-30}$  for $\theta = 15^0$$^{(9)}$. Then the 
second term  of the above expression will be much larger than the first 
term, and effectively, Doppler broadenning can be neglected in comparison to 
that due to multiple scattering effect.
\vskip5pt
\noindent
Now if we consider the other condition, $ i.e.,~~\al'~<~|\beta|$, the series in
(22) will be a divergent one and $\dl_{N+1}^2$ will be finitely large for large
but finite $N$. However, if the condition (23) is to be satisfied, then the
shift in 
frequency will be larger than the width of the spectral lines. In that case
 the
condition $\al'~<~|\beta|$ indicates that blueshift may also be observed but
the width of the spectral lines can be large enough depending on how large the
number
of collisions is. So in general, the blueshifted lines should be of larger widths
than redshifted lines and may not be as easily observable.
\vskip10pt
\subsection{Effect of Source Frequency on Broadening}
\vskip5pt
\noindent
Rearranging  equation (24) we get,
\beq
\left(\dl_n-\frac{\om_0}{2}\right)^2 \ll
\frac{\om_0^2}{4}-\frac{\al'}{\al\al'-\bt^2} 
\eeq
Since the left side is non-negative, the right side must be positive. Moreover,
since the mean frequency of any source is always positive, {\it i.e.,}
$\om_0 \geq 0$, we must have
\beq
\om_0 \gg \sqrt{\frac{4\al'}{\al\al'-\bt^2}}.
\eeq
We take the right side of this inequality to be the {\bf \it critical source
frequency } ${\bf \om_c}$ which is defined here as.
\beq
\om_c = \sqrt{\frac{4\al'}{\al\al'-\bt^2}}.
\eeq
Thus for a particular medium between the source and the observer, the  critical
 source frequency is the lower limit of the frequency of any
source whose spectrum can be clearly analyzed. In other words, the shift of any
spectral line from a source with frequency less than the critical source frequency for that
particular medium cannot be detected due to its high broadening.
\vskip5pt
\noindent
 We now can classify the spectra of the different sources,  from which light 
comes
to us after passing through a scattering medium characterized by the parameters 
$\al$, $\al'$,
$\bt$. If we allow only small angle scattering in order to get prominent
spectra, according to Wolf mechanism,  they will  either be blueshifted or
redshifted. The redshift of spectral lines may or may not be detected according 
to whether
the condition (24) does or does not hold. In this way those sources whose spectra
are redshifted, are classified in two cases, {\it viz.,}  ${\bf \om_0~~>~~\om_c}$  and
{\bf $\om_0~~\leq~~\om_c$}. In the first case, the shifts of the spectral lines can
be easily detected due to condition (23). But in the later case, the spectra
will suffer from the  resultant blurring.
\vskip10pt
\subsection{Doppler Shift vs. Wolf Shift}
\vskip5pt
\noindent
As we have seen in section(3.1) and section(3.2), the mean frequency and width 
of a
spectral line change after each scattering. The changes are given by the
recurrence relations:
$$ 
\delta_{n+1}^2 =  \frac{1}{\alpha'}+\frac{\beta^2}{\alpha'^2}\delta_n^2;$$
and
$$\omega_{n+1} = 
\frac{\omega_n|\beta|}{\alpha'+(\alpha\alpha'-\beta^2)\delta_n^2}.$$
Using these, we can easily find the Wolf-contribution to the observed shift
assuming only that the source under consideration is quasi-monochromatic. For
this we rewrite the equation (21) as :
$$\delta_{n}^2 = \frac{\alpha'^2}{\beta^2}\delta_{n+1}^2- \frac{\alpha'}{\beta^2}.$$

Now, according to Schr\"{o}dinger $^{(13)}$, the width of the spectral lines must increase
as a result of collisional processes. If this is the sole reason for the broadening,
then we can estimate N, the number of collisions it undergoes in its way to
us by comparing the
width at each scatterer with a preassumed small $\delta$ [ the spectral width
of the quasi-monochromatic source ]. We can easily calculate
\beq
z_W=(1+z_{scat})^N-1
\eeq
where $z_{scat}$ is the $z$-number due to small angle scattering at each
scatterer.
The following graph (Fig.4)illustrates the Wolf contribution in various observed
shifts. 
\vskip10pt
\noindent
The solid line curve represents the Wolf-contribution, while the dotted line 
( y=x ) represents the observed $z$-number. Thus the difference in height of
these two curves provides the contribution due to the Doppler effect which 
up to  the
present day was assumed to be the total contribution to $z$.
\section { \bf Width of the spectral lines and the Tully-Fisher Relation}
\vskip5pt
\noindent
According to our result(13), we conclude that both the shift and the width
of a spectral line depend on the medium parameters. In Fig.5, we
show the variation of the spectral line width as a function of shift.
 We can write the relation between the width and the shift as
\begin{equation}
W = K {( a^2 + \delta_0^2 z^2 )}^{\frac{1}{2(1+z)}}
\end{equation}
\vskip5pt
\noindent
where $K$ is a constant, $a^2$ is the minimum broadening and $\delta_0^2$ is the
 spectral width
inherent to the source.
The above relation is valid for $ z > - 1$.
Taking the logarithm of both  sides we can write

\begin{equation}
ln W = \frac{1}{2(1+z)} ln K ( a^2 + \delta_0^2 {z}^2)
\end{equation}

It is well known $^{(14)}$ that the distance modulus $d$  can  be written in terms of
 $z$ as

\begin{equation}
 d = m - M =  42.38 - 5 log (\frac{H}{100}) + 5 log z + (1.086)(1-q_0)z
+ O(z^2)
\end{equation}
Where $H$ is the Hubble constant in km/sec/Mpc and $M$ refer to the absolute
magnitude.
Now for small z i.e. $ z <<1$ ,

\begin{equation}
d = 42.38 - 5 log(\frac{H}{100}) + \frac{1.086}{2} z \ \ \ {\rm taking} \ \ \
 q_0 = \frac{1}{2}
\end{equation}
\vskip5pt
\noindent
After simplification we can write
$$ z = \frac{d- C}{.543} ; \ \ \ {\rm where} \ \ \  C = 42.38 - 
5log(\frac{H}{100})$$
\vskip5pt
\noindent
Now substituting this value of $z$ in  equation(30) we can write
\begin{equation}
d = - \frac{1.086}{logE} log W + N
\end{equation}
\noindent
where $$ log E = log (K a^2) \  \ {\rm and} \ \ N = C + 0.543$$
\vskip5pt
\noindent
The above relation between the distance modulus and the width has
a striking similarity to the Tully-Fisher relation but without any angular
dependence. The reason is obvious since we have considered the shift and width
due to scatterings only without considering any rotational effects. 
It appears that in the case of photons which are emitted 
perpendicular to 
 the plane of
the galaxy, we will be observing those photons only without any rotational
effects.
It should be mentioned that Bottenelli et al.$^{(4)}$ considered linear turbulence
correction for the profile width as

\begin{equation}
W_{20,i,c} = \frac{[ W_{20,{\rm obsv}} - W_{t,20}]}{\sin i}
\end{equation}
\vskip5pt
\noindent
where $W_{20,{\rm obsv}}$ is the observed (profile) width at 20 percent peak maximum
, corrected for instrumental broadening, $W_{(t,20)}$ is the correction term for turbulence
and the factor $\sin i$ corrects for disk inclination.

It would be interesting to compare this contribution  for galaxy spectra
due to non-rotational
part with the width calculated from miltiple scattering theory.
These will be considered in a subsequent publication.
\vskip10pt
\section {\bf Possible Implications}
\vskip5pt
\noindent
Unlike the Doppler effect where the width of a line is unrelated to the shift,
 the Wolf mechanism predicts a tight relationship between the width and the
shift of a line. As such, it is evident from the above analysis that {\it Dynamic Multiple Scattering
Theory} within Wolf's framework might play a significant role for QSOs in the
following ways.
\vskip10pt
\begin{enumerate}

\item  The width  and shifts of various spectral lines in different regions 
, say for UV, optical etc.,for AGNs and
Quasars may be explained in a consistent manner if the information regarding
the environments around these objects were available.

\item  Since the Wolf contribution becomes prominent for larger $z$ as 
can be seen from
Fig.2, this mechanism may shed new light in explaining the redshifts of quasars 
which tend to be large and peaking at $z \sim 2-3$ as
well as for hight redshift Galaxies ($z > 1$).

\item  For blue shift (i.e., $\alpha' < |\beta|$ ), as the broadenning is
 much larger, as is evident from (22), than the shift, one would  get 
a continuous spectrum  with 
no discrete lines.

\item  For low redshifts we can derive a relation like the  
Tully-Fisher as a function for
distance modulus and width given by a physical mechanism. We also derived
a general relation for higher redshifts(eqn. (30)) which 
can be verified from  observations. Further studies can be made regarding the theoretical basis
of the Baldwin Effect $^{(15)}$ for line and continuum correlations in AGN.
\end{enumerate}
\vskip5pt
\noindent
Lastly, we should mention that the critical source frequency relating 
to the screening effect which plays a crucial role in explaining both  redhsift and
spectral width can be tested in {\it Laboratory Experiments}. 

\vskip40pt

{\bf Acknowledgements:} One of the authors ( S.D.) greatly acknowledges World
Laboratory, Laussane for financial support during this work and Prof. B. K.
Datta ( Director, World Laboratory, Calcutta Branch ) for encouragement. 
The author(S.R.) is indebted to Prof. Jack Sulentic,University of Alabama,
for valuable suggestions and comments.
\newpage
\vskip20pt

{\bf REFERENCES}

\begin{enumerate}

\item  Tully, R.B. and Fisher, J.R.(1977), A \& A, {\bf 54}, 661.
\item  Aaronson, M.,Huchra, J.P.,\& Mould, J.R.(1979), ApJ,{\bf 229},1.
\item  Matthews, L.D., Driel,van Driel \& Gallaagher,III J.S.(1998), 
       ``An Exploration of the Tully-Fisher for Extreme Late-Type Spiral 
       Galaxies'', Astro-ph/9810042.
\item  Roberts, M.S., AJ, (1978),{\bf 83},1026. 
\item  Bottenelli, L.,Gouguenheim, L., Paturel,G.,de Vaucouleurs,G.
       (1983),A \& A ,
       {\bf 118}, 4.
\item  Rhee,M.H. (1996),Ph.D. Thesis, University of Groningen.
\item  Datta S, Roy S, Roy M \& Moles M, (1998), Int. Jour. of Theo. Phys., 
       {\bf 37},N4, 1313.  
\item  Datta S, Roy S, Roy M \& Moles M, (1998), Int.Jour.Theort.Phys., 
      {\bf 37},N5, 1469.
\item  Wolf E. and James D.F.V.(1996),Rep. Progr. Phys.{\bf 59}, 771.
\item  James D.F.V and Wolf E, (1990), Phys. Lett. A, {\bf 146}, 167.
\item  Born, M.\& Wolf,E. (1998), ``Principle of Optics'', 6th edition, 
       Pergamon, Oxford.
\item  Datta S, Roy S, Roy M and Moles M, (1998), Phys.Rev.A,{\bf 58},720.
\item  Schr\"{o}dinger E., (1955), IL Nuovo Cimento, {\bf 1}, 63.
\item  Narlikar, J. V., (1983),  ``Introduction to Cosmology'', Jones \& 
       Bertlett Publishers, Inc., Boston.
\item  Osmer, S.Patrick, Shields C.Joshep ,(1998), ``A Review of Line and Continuum
       Relations in AGNs'', astro-ph/9811459.
 
\end{enumerate}

\newpage

{\bf Figure Captions :}

\vskip10pt
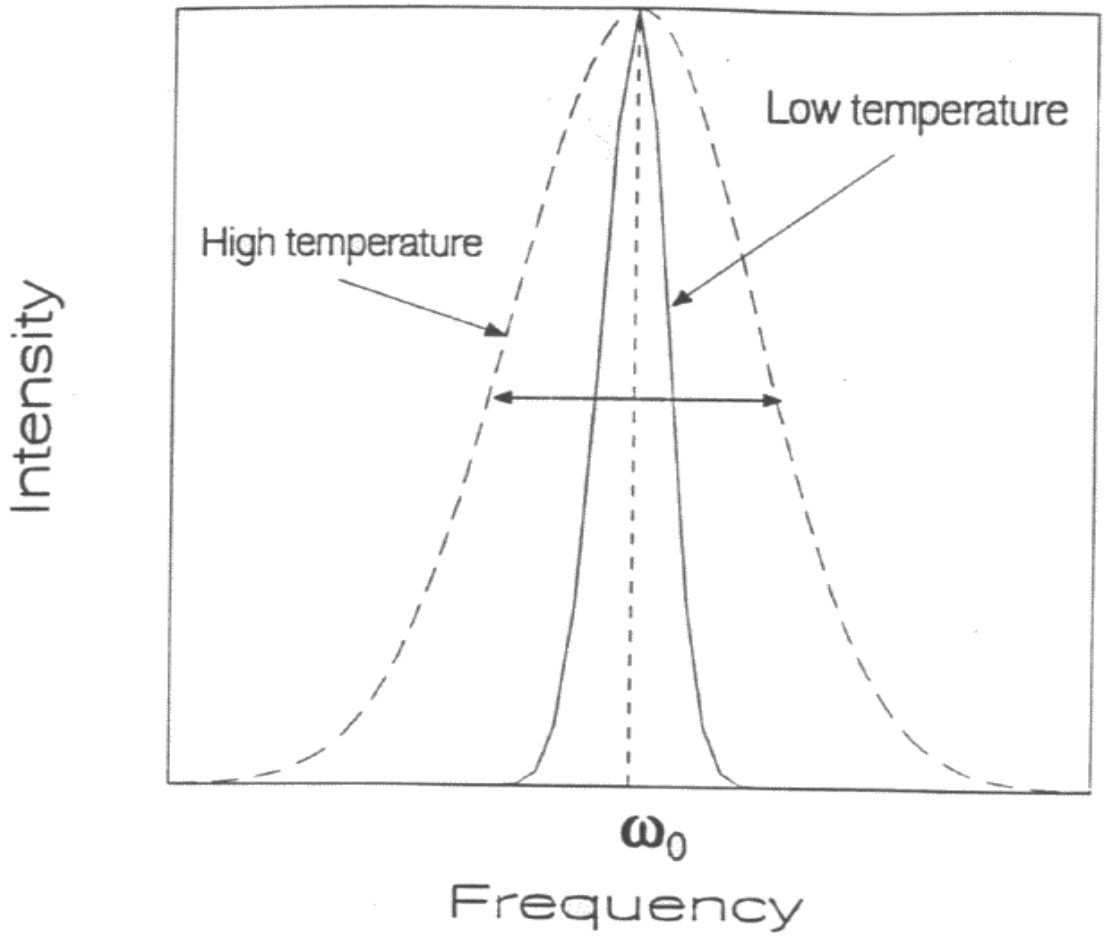
Figure 1 : Doppler Broadening of Spectral Line.
\vskip10pt
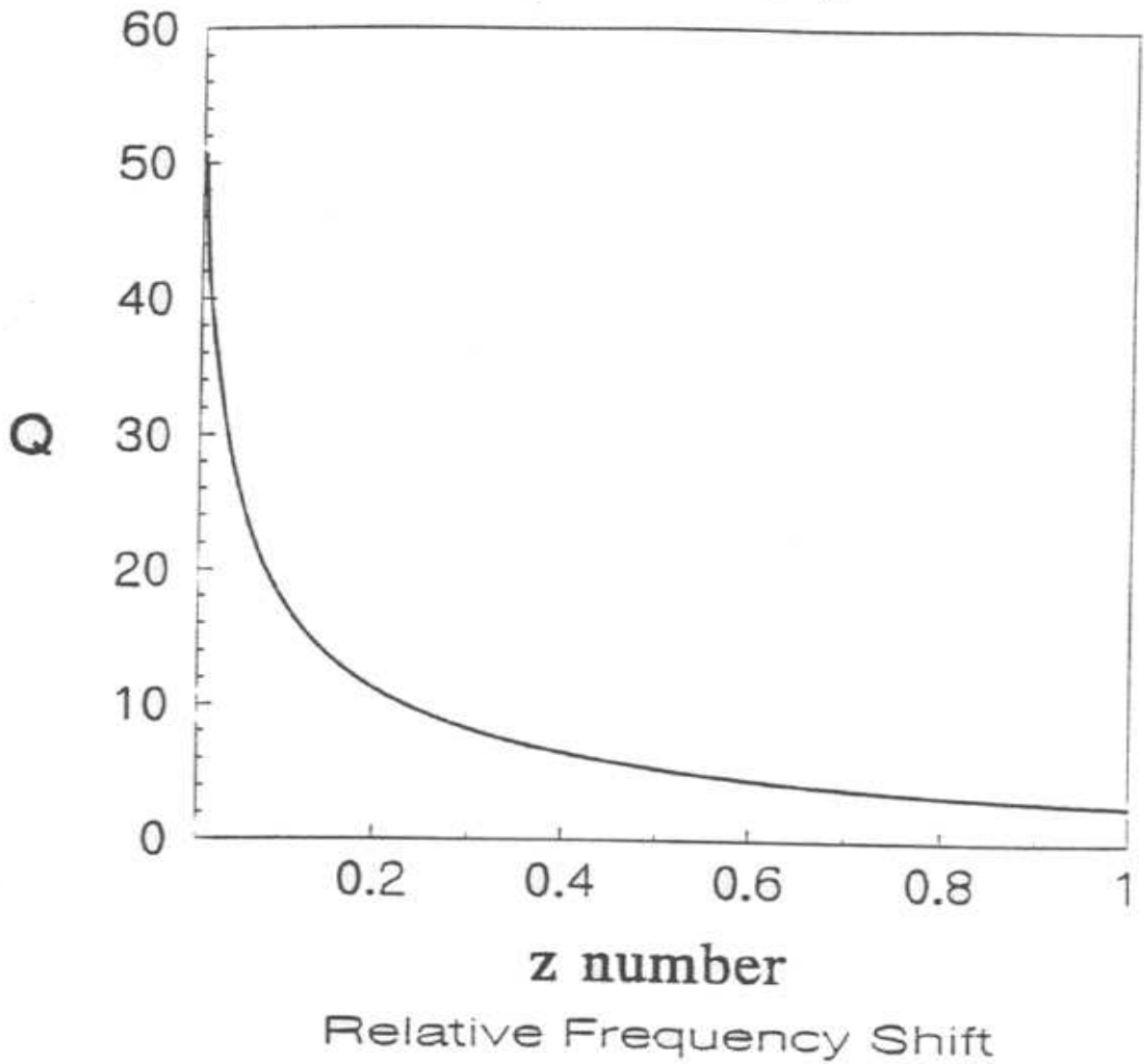
Figure 2 :Variation of Sharpness of a Spectral Line with Relative Frequency Shift
(for low $z$).
\vskip10pt
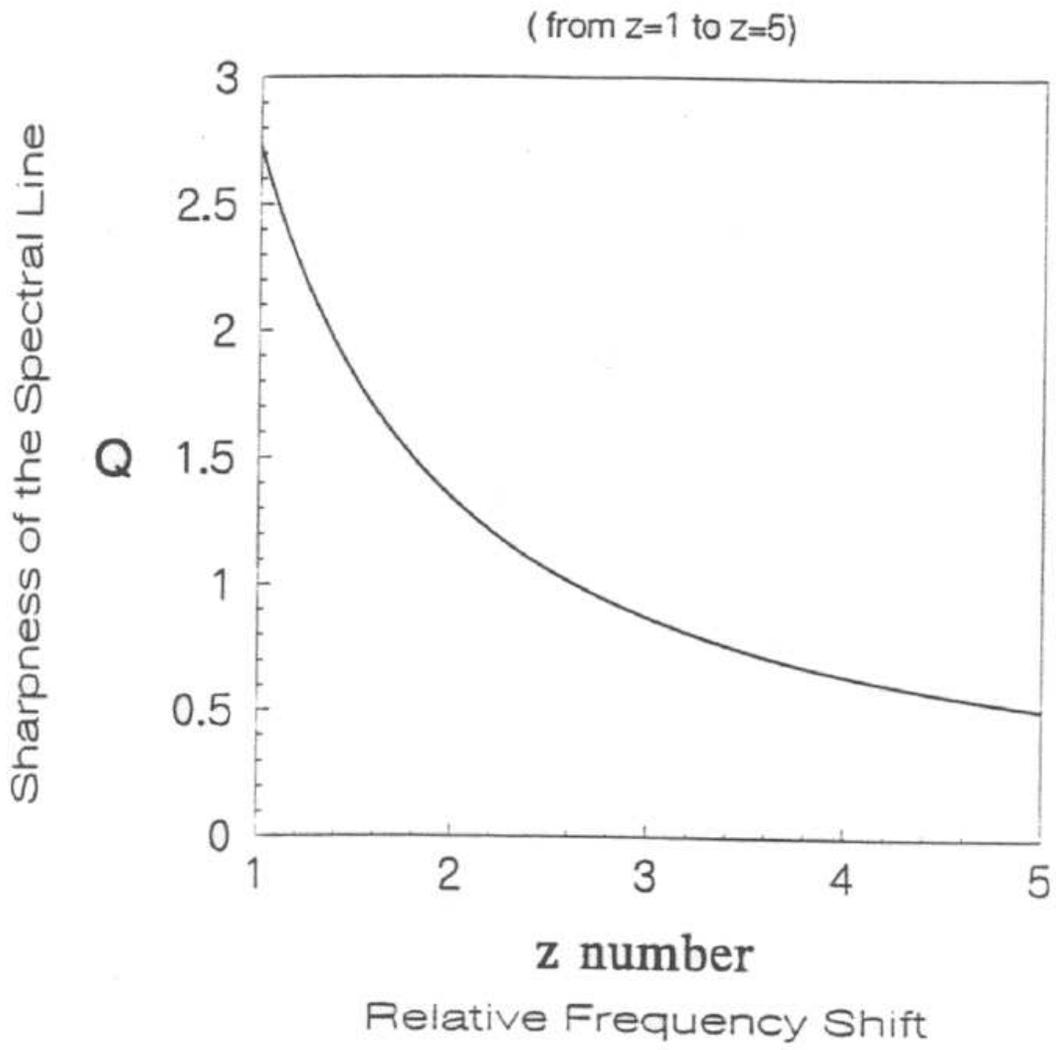
Figure 3 : Variation of Sharpness for high ($z$).
\vskip10pt
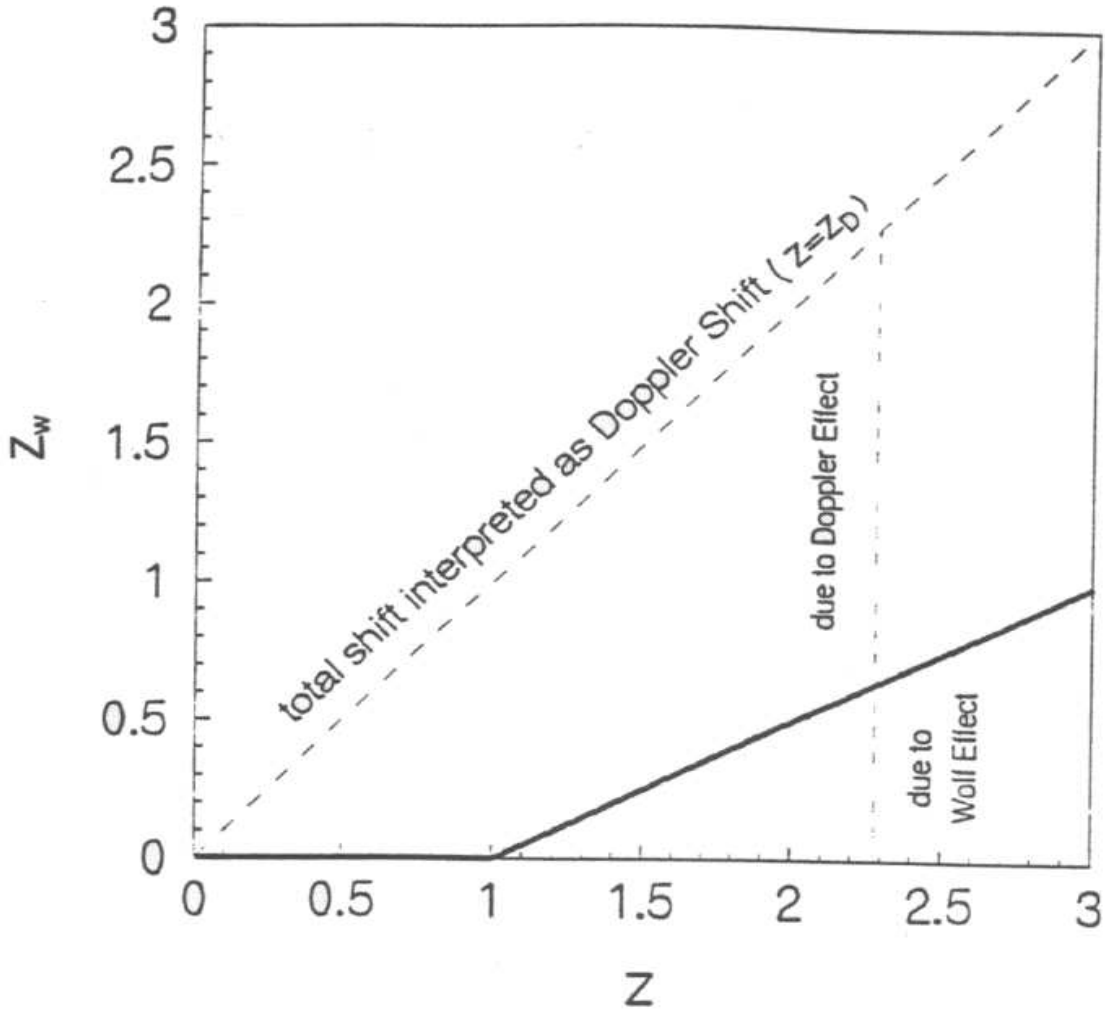
Figure 4.Wolf Contribution in Relative Frequency Shift
\vskip10pt
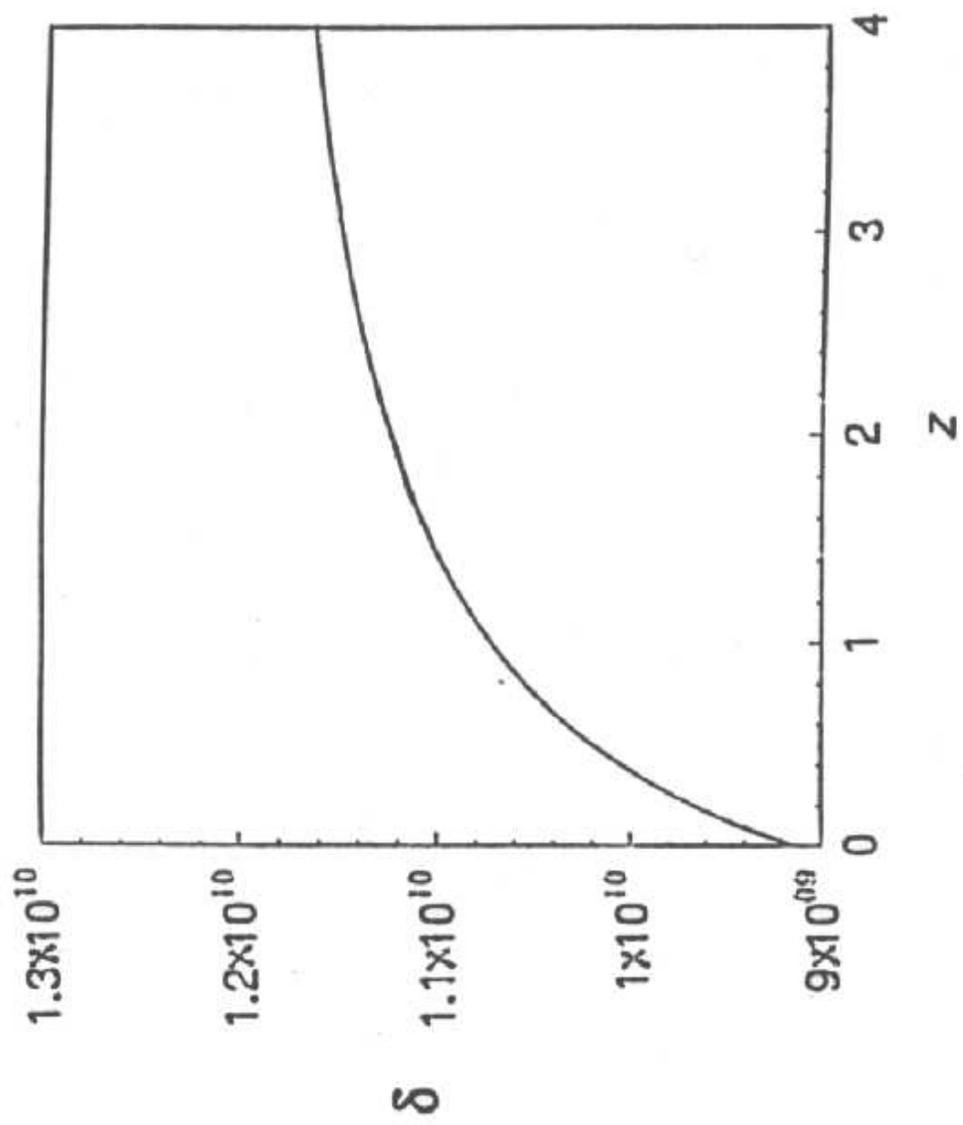
Figure 5. Variation of Width with shift.

\newpage

{\bf Summary of the changes and response to all recommendations and
criticisms} \\

\vskip10pt
\noindent
\begin{enumerate}
\item   On page 3 : we have given the refernce of E.Wolf and D.V.F.James:
        Rep.Prog.Phys. {\bf 59}, 771 (1996). 

\item   On page 6 : we have quoted the above reference as suggested by
        the referee.

\item   In eqn(8)  : we have rewritten the equation as per suggestion
         of the referee

\begin{equation}\displaystyle{
S^{(\infty)}(r \vec{u'} ,\omega')=A\omega'^4\int_{-\infty}^{\infty}
K(\omega,\omega',\vec{u}, \vec{u'})S^{(i)}(\omega)d\omega}
\end{equation}

\item   On page 8 : the refernce was given as James and Wolf instead of
        Wolf.

\end{enumerate}
\vskip10pt
\noindent
{\bf Response to Valerie Miller regarding figure problems with MSAA7127} \\
\vskip10pt
\noindent
 
All the necessary changes have been made as suggested by Valerie Miller.

\newpage
To \\
The Editor \\
Physical Review A \\
\vskip5pt
\noindent
{\bf Re : MS AA7127}
\vskip10pt
\noindent
Dear Sir, \\
        We are resubmitting our MS AA7127 ( Broadening of spectral 
lines.. by Roy, Kafatos and Dutta) considering all the recommendations
of the referee for your consideration. We are enclosing a summary of
the changes made in the text as well as the changes in the figures as
suggested by Valerie Miller seperately. Moreover, we have typed the
whole MS in double spaced style as suggested by the editor.
Hope it will suffice your purpose.

With best regards,

\hfill \ \                                   Yours Sincerely, \ \

 \hfill \ \                                   ( Sisir  Roy).

\vskip10pt
\noindent
Address for Communication : \\
Prof.Sisir Roy \\
Center for Earth Observing and Space Research \\
Institute of Computational Sciences and Informatics \\
George Mason University \\
Room 113, Science and Technology 1 \\
Fairfax, VA 22030-4444 \\
USA . \\
Fax : (703) 993 3628.
\end{document}